\documentclass[11pt,twoside]{article}
\usepackage{amsmath}
\usepackage{amssymb}
\usepackage{amsfonts}
\usepackage{wrapfig}
\usepackage{graphicx}
\usepackage[eqsecnum]{cmp2e}
\def\be{\begin{eqnarray}}
\def\ee{\end{eqnarray}}

\title[Effects of correlated disorder\ldots{}]%
{Effects of aerogel-like disorder on the critical behavior of $O(m)$--vector models.
Recent simulations and experimental evidences.
\thanks{Contribution to the {\em Mochima theoretical physics spring school}.
Joint CEA-IVIC-SFP workshop on Foundations of Statistical and
Mesoscopic Physics. \protect\\ [0.9ex] \strut\qquad 2005 World Year of Physics.}}
\author[V\'asquez and Paredes]{Carlos V\'asquez\refaddr{usb} and
Ricardo Paredes\refaddr{ivic,delft} }
\addresses{
\addr{usb} Departamento de F\'{\i}sica, Universidad Sim\'on Bol\'{\i}var,
Apartado 89000, Caracas 1080A, Venezuela.
\addr{ivic} Centro de F\'{\i}sica,
Instituto Venezolano de Investigaciones Cient{\'\i}ficas,
Apartado 21827, Caracas 1020A, Venezuela.
\addr{delft} Particle Technology Group, Delft University of Technology,
Jualianalaan 136, 2628 BL Delft, The Netherlands.
}

\begin{document}

\maketitle

\begin{abstract}
We review recent results on the effect of a specific type of quenched disorder on well
known $O(m)$--vector models in three dimensions: the {\em XY} model (3DXY, $m=2$)
and the Ising model (3DIS, $m=1$). Evidences of changes of criticality in both systems,
when confined in aerogel pores, are briefly referenced. The 3DXY model represents the
universality class to which the $\lambda$--transition of bulk superfluid $^4$He
belongs. Experiments report interesting changes of critical exponents for this
transition, when superfluid $^4$He is confined in aerogels. Numerical evidence has
also been presented that the 3DXY model, confined in aerogel-like structures, exhibits
critical exponents different from those of bulk, in agreement with experiments. Both
results seem to contradict Harris criterion: being the specific heat exponent negative
for the pure system $(\alpha_{\mbox{\tiny 3DXY}}\simeq -0.011<0)$, changes must be
explained in terms of the extended criterion due to Weinrib and Halperin, which
requires disorder to be long-range correlated (LRC) at all scales.
In numerical works, aerogels are simulated by the {\em diffusion limited cluster-cluster aggregation}
(DLCA) algorithm, known to mimic the geometric features of aerogels. These objects,
real or simulated, are fractal through some decades only, and present crossovers to
homogeneous regimes at finite scales, so the violation to Harris criterion persists.
The apparent violation has been explained in terms of {\em hidden} LRC subsets within
aerogels $[$Phys. Rev. Lett., 2003, {\bf 90}, 170602$]$. On the other hand, experiments
on the  liquid-vapor (LV) transition of $^4$He and N$_2$ confined in
aerogels, also showed  changes in critical-point exponents.
Being the LV critical-point in the $O(1)$ universality class,
criticality may be affected by both, short-range correlated (SRC) and LRC subsets of
disorder. Simulations of the 3DIS in DLCA aerogels can corroborate experimental
results. Experiments and simulations both suggest a shift in critical exponents to
values closer to the SRC instead of those of the LRC fixed point.

\pacs 64.60.Cn, 64.60.Fr, 64.70.-p

\end{abstract}

\section{Harris criterion in brief: original and extended}
Whether the presence of disordered impurities affects the critical behavior
of an ideal system, or not, has been the task of numerous works through
years. Since Harris' seminal work \cite{harris74}, a robust
theoretical background has emerged to establish conditions for the relevance
of disorder to phase transitions, which concern the criticality of the original
pure system as well as the geometrical features of the disordered distribution
of defects \cite{weinrib83,aharony96,korshii,prudnikov00}.
On the other hand, many experimental contributions in the last two decades posed
interesting questions about criteria of relevance,
supplying results that challenge predictions made by previous theoretical works.
Amongst all, those on the superfluid transition of $^4$He in light aerogels
are rather intriguing \cite{chan88,chan96,wong90,mulders91,larson92,yoon98}.

Changes of critical exponents for the so called $\lambda$--transition,
when superfluid $^4$He is confined in aerogel pores,
were reported repeatedly from late 80's through late 90's,
with almost the same question left open:
{\em do these results violate Harris and/or other relevance criteria?}
The question arises because aerogels are homogeneous (non-correlated) beyond
a finite scale \cite{vacher88,hasmy94}, and given that the specific heat exponent
($\alpha$) is negative for this transition, exponents should not change after
{\em Harris criterion}. Many numerical and theoretical works have since emerged to explain
these contradictory results \cite{korshii,prudnikov00,machta90,li90,moon95,vasquez03}.

According to Harris criterion \cite{harris74}, {\em short-range correlated} (SRC) disorder is
irrelevant for the critical behaviour of any $d$-dimensional pure system which undergoes a
second order phase transition with a correlation length exponent $\nu_{pure} > 2/d$.
The criterion was shown valid if disorder presents a correlation function
$\delta(\vec{\boldsymbol{r}})$ and, after Josephson hyperscaling\cite{joseph66},
$\alpha_{pure}<0$. This original work was improved several years later by Weinrib and
Halperin (WH)\cite{weinrib83}, who established a more general criterion of relevance:
even if $\alpha < 0$ for the pure system, it will change critical exponents if disorder
is ``correlated enough''. A proper definition of this {\em long-range correlated} (LRC) disorder
makes use of the {\em impurity-impurity correlation function} $g(r)=\langle n(r)n(0)\rangle$ and its
long-range scaling exponent ($g(r)\sim r^{-a}$, $r\to \infty$). Depending on how fast the
tail of $g(r)$ decays, the {\em extended} criterion reads as follows:
\begin{wrapfigure}{i}{0.5\textwidth}
\vspace{0.25cm}
\centerline{\includegraphics[width=0.5\textwidth]{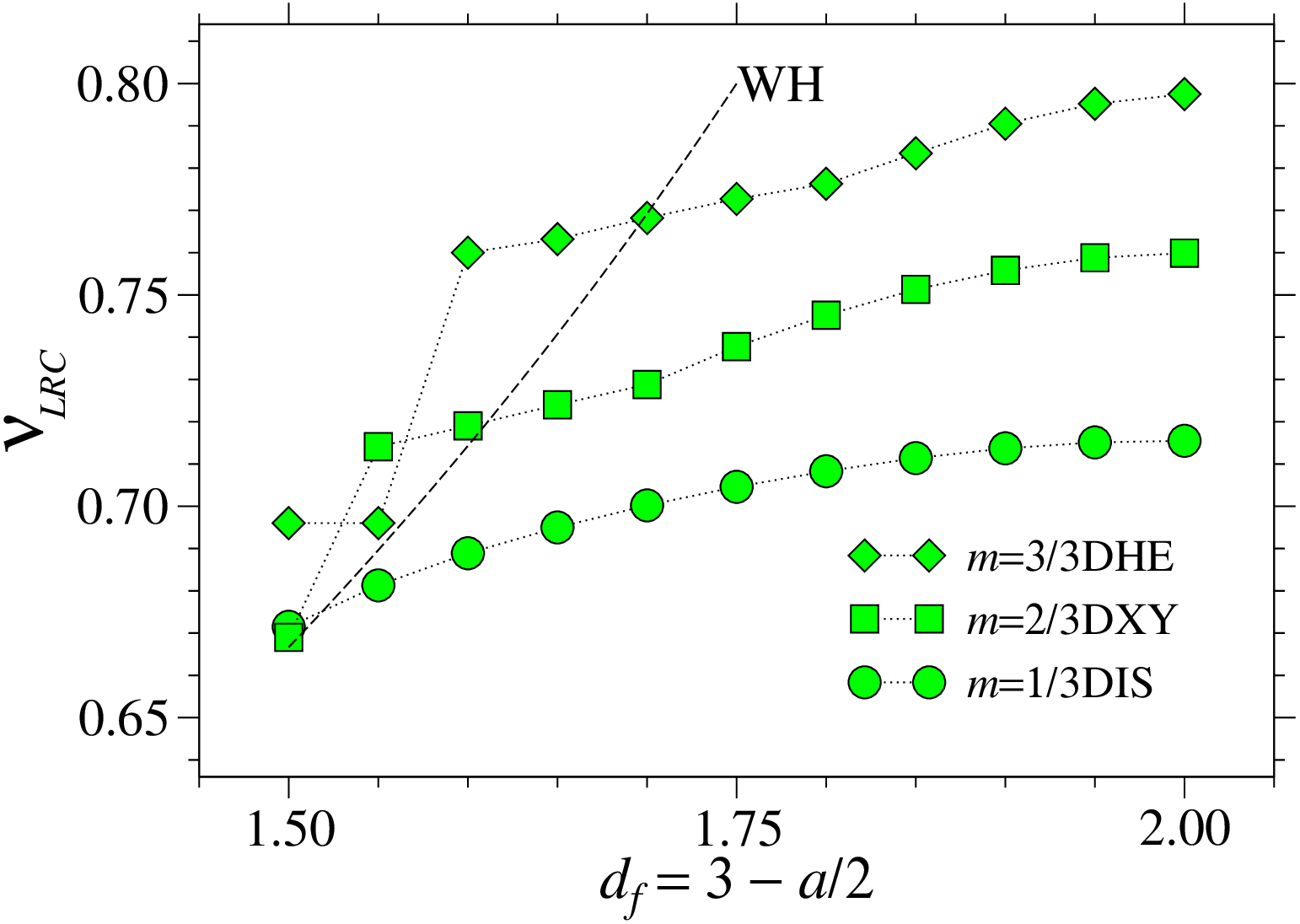}}
\parbox{0.55\textwidth}{
\caption[Prudnikov's exponents]
{\label{prud}{
Correlation length exponent for $O(m)$--vector models ($m\!=\!1,2,3$) with LRC defects,
depicted as functions of $d_f=d-a/2$. WH $m$-independent result plotted for comparison
(dashed line). Data from Table V in Ref. \cite{prudnikov00}.}}
}
\end{wrapfigure}
\be
2-d\nu_{pure}  \geq 0\; &\mbox{for}\;& a \geq d\label{whsrc}\\
2-a\nu_{pure}  > 0\; &\mbox{for}\;& a < d.\label{whlrc}
\ee
In both cases disorder is relevant. Case (\ref{whsrc}) extends the particular
definition of {\em short-range correlated} (SRC) disorder, made by Harris, to disorder
distributions with a ``fast-decaying'' tail ($a>d$). The second case (\ref{whlrc})
is valid even if $\nu_{pure} > 2/d$. Based on one--loop expansions in
$\epsilon=4-d\ll 1$ and $\delta=4-a\ll 1$, WH also argued that, the correlation length
exponent would be $\nu_{\mbox{\tiny LRC}}=2/a$, independent on the internal dimension
($m$) of the order parameter. Prudnikov {\em et~al.~}\cite{prudnikov00} performed more
accurate estimates to deduce $m$-dependent corrections, {\em i.e.}, under the same kind
of LRC distribution of defects (same $a$), their two-loop field theoretical expansions
give different critical exponents for different $O(m)$ systems.
Fig. \ref{prud} resumes graphically what they obtain for the correlation length
exponent $\nu_{\mbox{\tiny LRC}}$, at disorder correlation exponents $2<a<3$ and
$m=1,2,3$, quite different from the $m$--independent plot predicted by WH
\cite{weinrib83}.

\section{Phase transitions in aerogels: experiments}
In this section, we review some important experimental results about
critical systems confined in aerogels. In $1988$, Moses Chan {\em et~al.~}\cite{chan88}
reported on the influence of quenched disorder on the superfluid (SF)
transition of confined $^4$He, using three types of porous glass:
Vycor, aerogels and xerogels. They measured the temperature dependence
of the relative superfluid density, which scales as $\rho_s/\rho\sim t^\zeta$
near the critical point, being $t=|T-T_c|/T_c$ the reduced temperature,
and the exponent $\zeta\equiv\nu$ due to hyperuniversality \cite{mulders91,yoon98}.
In the first case ($^4$He--Vycor), they observed no change in the critical exponent
from $\zeta\simeq 0.6705$ of bulk SF $^4$He, a fact clearly explained in terms of
Harris criterion, provided that the pure critical system exhibits a negative
specific heat exponent ($\alpha_{pure}\simeq -0.0105$) \cite{lipa96}.
Internal microchannel structures of Vycor, randomly oriented and randomly distributed,
present scattering intensities that exhibit a peak near $r_{max}\sim 20$nm,
and fall off exponentially above this peak, clearly faster than $r^{-d}$.
In terms of the WH model, this $^4$He--Vycor system falls into the SRC regime
(\ref{whsrc}), so theoretical predictions agree with experiments.
Alternatively, Zassenhaus and Reppy \cite{zassen99}, reported calorimetric studies
for this $^4$He--Vycor system, showing that the singular part of specific heat curves
fit well a $\lambda$--like curve with the same exponent $\alpha$ as bulk SF $^4$He.
Further experiments, using porous gold to confine SF $^4$He, instead of Vycor,
confirm these results \cite{yoon97}. Microchannels within these structures
are similar to those of Vycor, but porosities are greater.

The challenge to theory comes in the case of $^4$He confined in silica aerogels (AE)
and xerogels (XE), where authors observed
$\zeta_{\mbox{\tiny AE}}\simeq 0.81$ and $\zeta_{\mbox{\tiny XE}}\simeq 0.89$,
respectively, larger than the bulk exponent. After the WH model, these results suggest
that LRC distributions of defects must be present within AE and XE. In practice,
correlation functions are measured by means of (neutron$/$X-ray) scattered intensities
\cite{vacher88}, thus algebraic (power-law) regimes on these plots correspond to
algebraic regimes on $g(r)$. At these regimes, objects present self-similar (fractal)
structure. Thus, provided that the condition (\ref{whlrc}) is fulfilled by the exponent
($a$), critical exponents of the system will be modified by disorder.

Nevertheless, authors argue that silica AE certainly exhibit fractal regimes for several
length scales, but they point out that beyond a finite cutoff $\Lambda_{\mbox{\tiny AE}}$
curves enter homogeneous SRC regimes. Thus, close enough to the critical point, where
the correlation length of SF $^4$He diverges ($\xi\sim t^{-\nu}$), disorder may appear
homogeneous at the typical length scale of the system, and exponents must cross over to
bulk values. For instance, in the case of $95\%$ porous aerogels
 $\Lambda_{\mbox{\tiny AE}}\simeq 150$nm, and authors estimate $\xi(t)\simeq 480$nm
already at $t\simeq 10^{-4}$, but although they run experiments up to
$t\simeq 10^{-5}$, the crossover never appears\cite{yoon98}. They actually rule out
an apparent violation to Harris criterion, but leave the open question about the LRC
character of disorder, and propose the existence of LRC within aerogels, that cannot be
observed through conventional techniques \cite{chan96}.

An attempt to explain these changes on the critical behaviour of SF $^4$He will be
discussed below, in Section \ref{dlca}, showing that well defined LRC structures actually
exist within aerogels. In the XE case, instead, a clear explanation to changes in critical
exponents does not yet exist to date. As aerogels, these structures are created through
silica sol-gel aggregation, but microstructures (strands) are broken and reorganized at the
drying stage of the process, resulting in more compact (less porous) materials. It has been
suggested\cite{gimel95} that, close to the flocculation-percolation transition in the
aggregation process, strands present the same fractal dimension as the LRC structure,
but comparative studies do not yet reveal self-similarity in xerogels\cite{lattuada01}.

On the other hand, the  liquid-vapor (LV) critical point, known to belong to the 3DIS
universality class \cite{pestak84}, gives another scope into the influence of aerogel-like
disorder on phase transitions. In this case, {\em any} type of disorder, correlated or not,
is relevant for criticality of the pure system.
After Harris criterion \cite{harris74}, a positive specific heat exponent for the pure
system ($\alpha_{\mbox{\tiny 3DIS}}\simeq 0.109$) makes random SRC disorder already
relevant. In addition, theory predicts \cite{weinrib83,prudnikov00}, that the $O(1)$
universality class is also subject to changes in criticality under a LRC distribution
of defects.

As we quote \cite{hasmy94,vasquez03} and show below, both SRC and LRC structures are
present within aerogels. Thus, the 3DIS confined in aerogel-like distributions of defects
(AEIS) may be subject to these two competing effects. Results presented by Wong
{\em et al.~}\cite{awong90}, show that the LV critical point of fluid $^4$He in aerogels
exhibits specific heat curves with a finite peak at $T_c$, which suggests that
$\alpha_{\mbox{\tiny AEIS}}<0$, clearly different from the pure system exponent.
In these and further experiments, using N$_2$ instead\cite{awong93}, authors also report on
the order parameter exponent concluding that, within error bars, it is indistinguishable
from the pure system exponent. Theoretical predictions on the 3DIS with defects, give
second order corrections to magnetic exponents respect to those of the pure system.
In addition, the coexistence curve is about $12$ times narrower than that of the pure
system \cite{awong90}, so it is not surprising that changes were so difficult to detect.

The only trace then left of an effect of aerogel-like disorder on the LV critical point is
that concerning the finite peak of the specific heat. In a recent paper\cite{paredes},
Paredes and V\'asquez show that this finite peak is consitent with a shift in the
correlation length exponent to a value close to $2/3$, which is closer to the corresponding
exponent for the randomly diluted Ising system (RDIS)
\cite{holovatch,calabrese,pelissetto}, which is SRC at concentrations of disorder $c$ below
the percolation critical value $p_c$. However, it is hard to reach definitive asymptotic
values for the exponents, due to an apparent oscillating approach to the fixed point.
The same question emerges thus from both, experiments and simulations:
{\em why criticality would shift to the SRC and not to the LRC fixed point?}

\section{Correlations within aerogel-like distributions of disorder: DLCA}\label{dlca}
In the sol-gel process of aerogel construction, silica dust is suspended in a solvent
({\em sol} phase), which allows diffusion and cluster-cluster aggregation to take place.
Once the process has finished and the {\em gel} has been dryed, detailed analyses of
structure reveal that these objects are neither fractal nor LRC in the WH sense
(\ref{whlrc}): only finite regions of power-law scaling are observed in
scattering intensities from these objects \cite{vacher88}.
However, at a given time $t_g$ of the process, one of the clusters spans all the space
of the flask, the {\em gelling cluster} (GC).
Although it has not been tested experimentally, evidence exists that these GC
are fractal at all scales \cite{hasmy94} and LRC in the WH sense \cite{vasquez03}.
A numerical tool, the {\em diffusion limited cluster-cluster aggregation algorithm} (DLCA),
developed in early 80's separately by Meakin\cite{meakin} and Kolb, Botet, and Jullien
\cite{remiju}, has proved to reproduce well the geometrical features of real aerogels
\cite{hasmy94}.

As silica dust in suspension, in DLCA, particles that initially occupy random positions
in a 3D box, are allowed to perform independent Brownian motions, and then kept
together whenever two of them enter in contact.
Undergoing this process, monomers and aggregates continue to move randomly,
and attach themselves to each other until a unique cluster has finally been formed.
In the on-lattice version of the algorithm, under periodic boundary conditions,
$cL^3$ particles occupy random sites in a simple cubic $L^3$ box, being $c$ the
concentration and $\varphi=1-c$ the porosity. A mass dependent diffusive constant is
taken into account through a probability $p\sim m^{-\varsigma}$ for an aggregate of
mass $m$ to move, where $\varsigma$ is a positive mobility exponent.

As in experiments, the GC is defined as the first cluster to reach opposites sides of
the box in any direction. Right after the GC is built, many other smaller clusters
({\em islands}) continue to diffuse, and finally attach themselves at random sites to the
GC. In the fabrication of an aerogel, the resulting structure is preserved by
hypercritical evaporation of the solvent \cite{vacher88}, so the GC matrix is kept
almost untouched, with randomly distributed strands attached to it. Islands represent
the SRC subset within the whole cluster.
\begin{wrapfigure}{i}{0.5\textwidth}
\vspace{0.2cm}
\centerline{\includegraphics[width=0.5\textwidth]{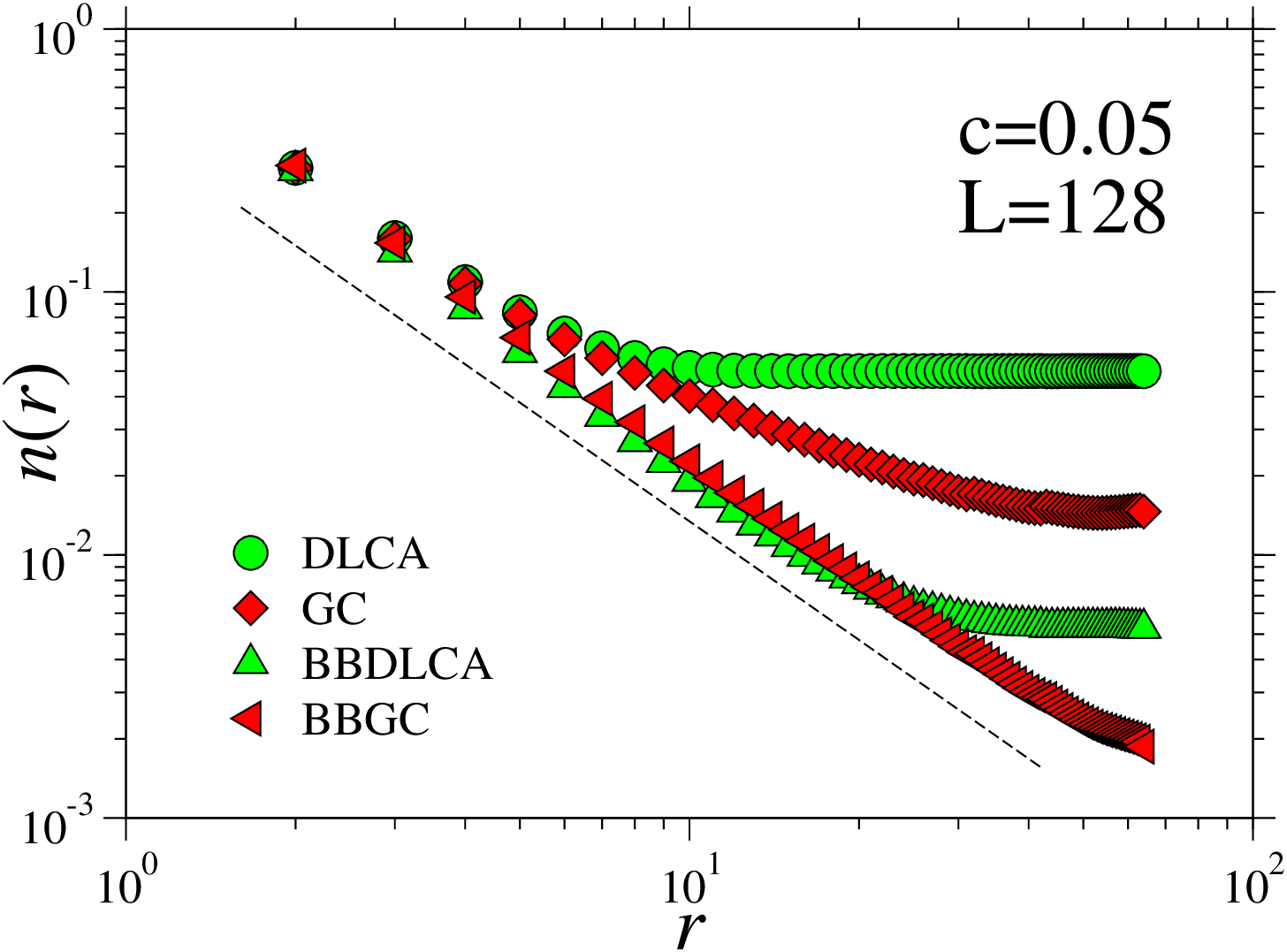}}
\parbox{0.52\textwidth}{
\caption[Correlations in DLCA]
{\label{sub-dlca}{Average density $n(r)$ versus box size $r<L/2$ for different subsets of
$c=0.05$ DLCA clusters ($L=128$). The dashed line represents the slope $a/2=1.5$. }}
}
\end{wrapfigure}

Fig.~\ref{sub-dlca} shows the geometrical features of different subsets of DLCA clusters,
built in $L=128$ boxes at concentration $c=0.05$. Following a box-counting procedure
\cite{feder}, the scaling of corresponding average densities $n(r)=m(r)/r^d$ have been
depicted as function of the box size $r$. Average over $32$ realizations is taken.
Plots marked as BBDLCA and BBGC correspond to {\em backbones} for the complete DLCA structure
and the GC, respectively. Plots corresponding to complete DLCA clusters (circles) fast
enter to the homogeneous regime, due to islands attached to random sites at gelling
clusters (squares): the slow power-law decay of GC alone suggest that these objects are
LRC, but when islands have already reached the GC to finally form the whole DLCA structure,
the LRC regime disappears screened ({\em hidden}) by the homogeneous distribution of islands.
Backbones BBDLCA and BBGC also exhibit power-law long regimes, but as we argue just
below, exponents larger than $3/2$ give $a>d=3$, which make these subsets enter the SRC
class, after WH criterion (\ref{whsrc}).

The average mass $m(r)$ scales as $r^{d_f}$, being $d_f$ the fractal dimension of the
object. Recall that in the box-counting procedure one only takes non-empty boxes to
estimate the average. This is equivalent to consider $n(r)$ as a {\em one-point} correlation
function: considering only occupied sites as centers for $d$-dimensional cubes, the
algorithm effectively measures the probability for surrounding sites to be occupied,
normalized to $r^d$. This method rends a decaying exponent $(d-d_f)$ for $n(r)$. We have
argued\cite{vasquez03} that the decaying exponent for this one-point correlation function
is related to that of the impurity-impurity correlation function $g(r)\sim r^{-a}$ as
$a= 2(d-d_f)$.

An example for this assertion comes from the percolation model, whose density-density
correlation function scales as $g(r)\sim r^{-(d-2+\eta)}=r^{-a}$ at the critical occupation
fraction $p=p_c$. Using Rushbrooke hyperscaling relation one obtains $d-2+\eta=2\beta/\nu$,
being $\beta$ the critical exponent for the order parameter ${\cal{P}}_{\infty}$. Remind
that the order parameter in the percolation problem is ${\cal{P}}_{\infty}=m_{\infty}/N$,
where $m_{\infty}$ is the mass (size) of the percolating cluster at $p_c$. This quantity
scales as $P_{\infty}\sim L^{-(d-d_f)}$, and being this the order parameter for the
percolation transition\cite{feder}, its exponent is defined as $d-d_f=\beta/\nu$. In this
case, it follows that the relation stated above, $a=2(d-d_f)$ seems to hold \cite{comment}.

\begin{figure}[!ht]
\vspace{0.25cm}
\begin{center}
\hspace{-0.5ex}\includegraphics[width=0.51\textwidth]{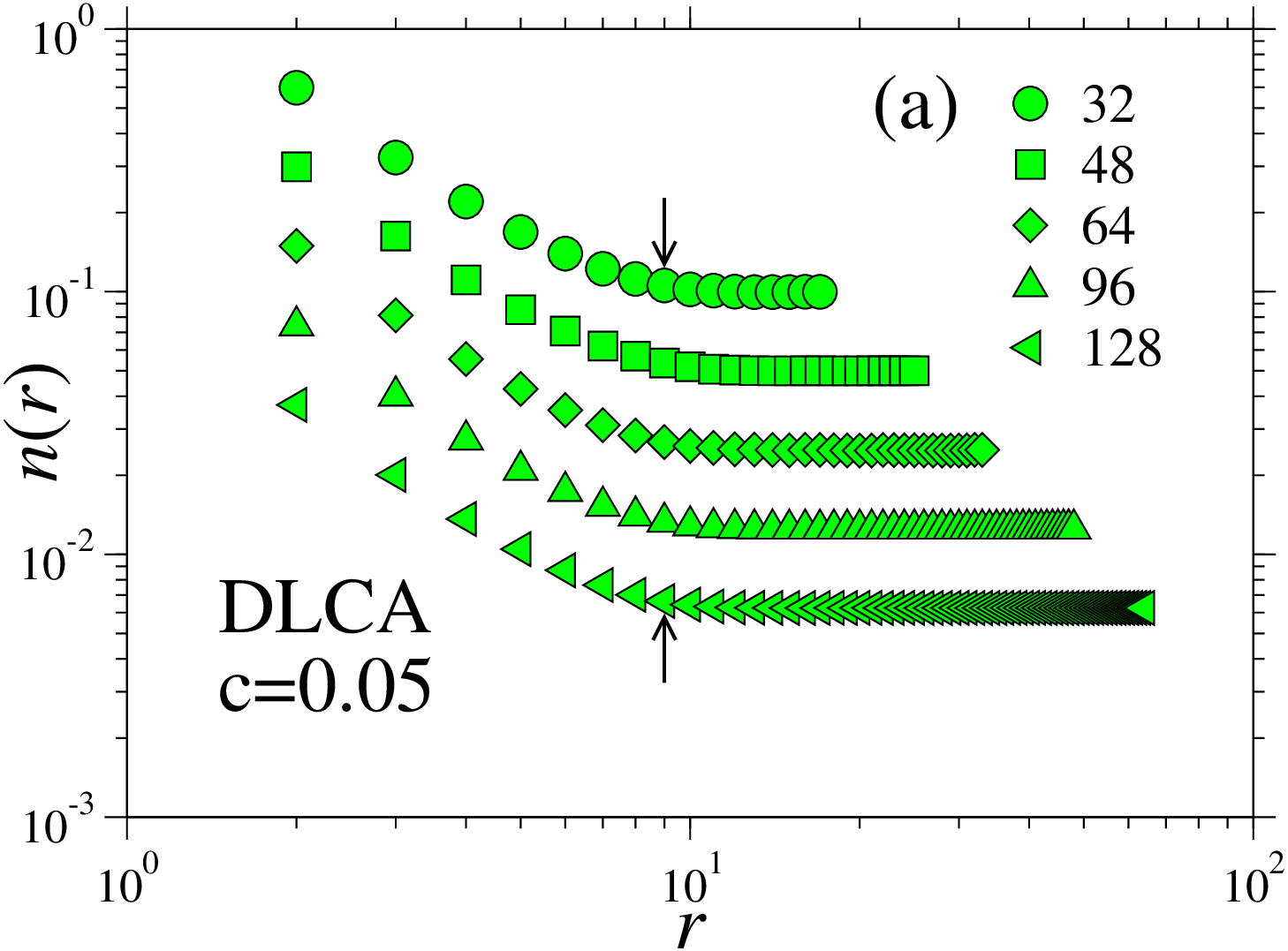}
\hspace{0.25ex}\includegraphics[width=0.475\textwidth]{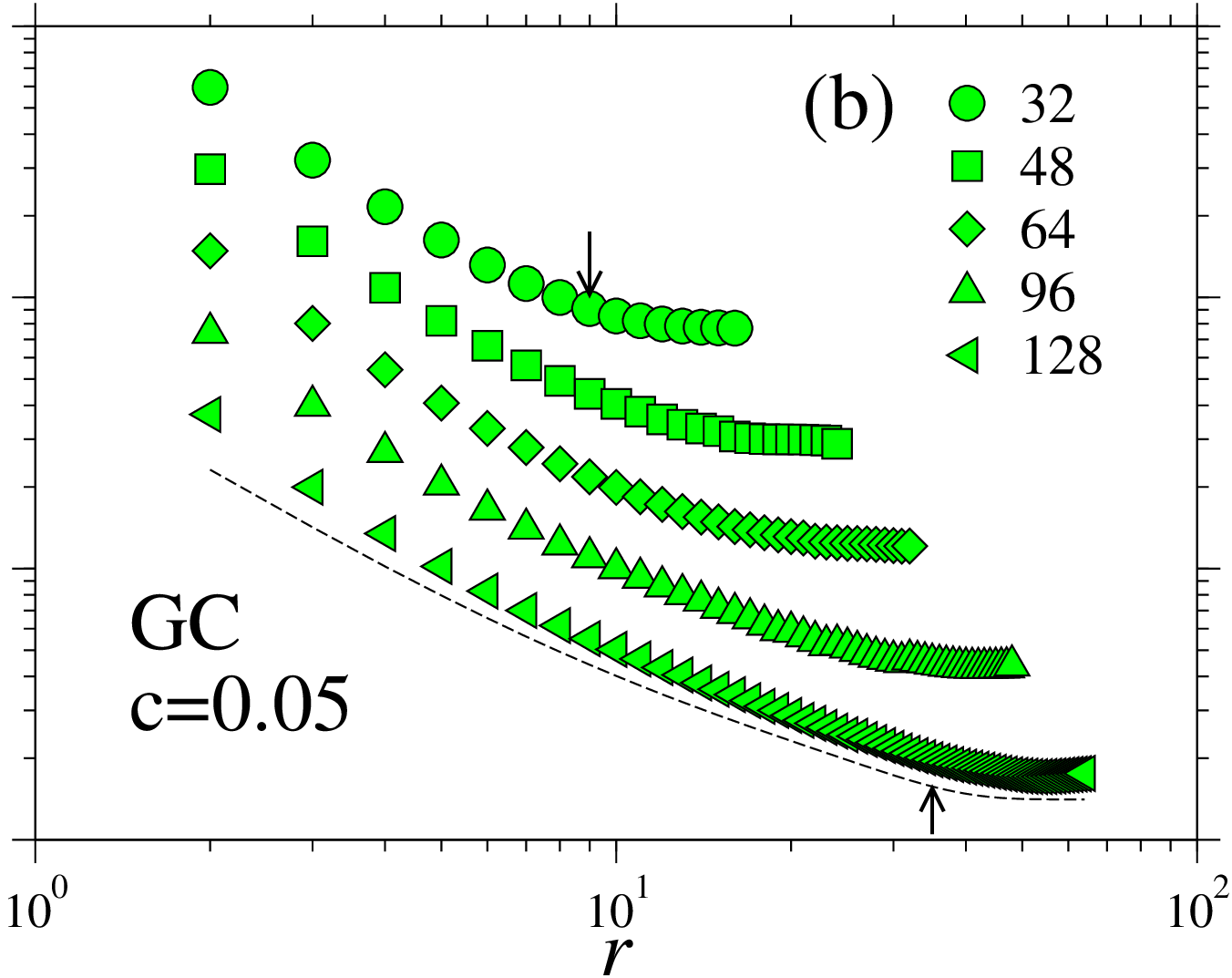}
\parbox{\textwidth}{
\caption[Correlations in DLCA-GC]
{\label{corr-dlca}{Average density $n(r)$ versus box size $r$ for (a) DLCA complete
clusters and (b) corresponding GCs. Crossover to homogeneous regimes indicated by
arrows. The dashed curve in (b) is the stretched exponential fit for the GC at $L=128$
(see text). Plots have been vertically displaced for better comprehension. }}
}
\end{center}
\end{figure}

For the complete DLCA cluster, the crossover to a homogeneous regime occurs at $r\ll L/2$,
while the fractal regime for backbones seems to span all scales. In the WH scheme, however,
backbones present the fast decaying SRC regime (\ref{whsrc}), comparing the corresponding
$n(r)$ with the slope $1.5$ (dashed line). The long-range decay for the GC seems slower
than $r^{-d}$, which enters in the LRC regime (\ref{whlrc}). This case is more complex as
it becomes homogeneous at $r\approx L/2$, probably because some strands are already
randomly attached to it at the gelation time $t_g$. To show that GC are in effect fractal
at the thermodynamical limit $L\to \infty$, and complete DLCA are not, plots of $n(r)$ for
different lattice sizes $L=32-128$ are shown in Fig. \ref{corr-dlca}.
Arrows indicate crossover lengths which, for DLCA clusters (a), reach
no more than a few lattice constants, independent of $L$. On the other hand,
arrows in Fig.~\ref{corr-dlca} (b) indicate that crossover lengths of GCs scale with the
lattice size, a trace of fractality of these objects at $L\to\infty$.
The complete DLCA cluster contains this LRC structure within, with numerous islands
randomly attached to it, which hide correlations in the box-counting numerical method and
make the whole structure appear homogeneous. This screening effect is unavoidable in
experimental measurement of correlations (small-angle scattered intensities), resulting in
finite fractal regimes for silica aerogels, which tend to disappear as density
increases\cite{vacher88}. For simulated aerogels, finite size modifies the long-range
behaviour of the correlation function. This effect is taken into account through the
stretched exponential expression proposed by authors in Refs.~\cite{gimel95,lach-hab98}
$n(r) = c_g + A_or^{-(d-d_f)}\exp[-(r/\xi)^{\delta}]$, where $c_g$ is the actual density of
the GC (which tends to zero as $L\to\infty$) and $\xi$ is the characteristic radius of
gyration (cuttoff). The parameter $\delta>1$ describes the faster crossover to the
homogeneous regime characteristic of finite lattices. The dashed curve in
Fig.~\ref{corr-dlca} (b) represents the fit to this expression for $L=128$, giving the
estimate $d_f\approx 1.7$ (thus $a\approx 2.6$) for the fractal dimension of the GC,
with a cutoff at $\xi\approx 35$.

\section{$O(m)$--vector models with aerogel-like distribution of defects}\label{AEXY}
Monte Carlo simulations of $O(m)$--vector models consider nearest neighbor interactions
between spins, placed on a 3D simple cubic lattice of size $L$ ($1<i<L^3$), with
periodical boundary conditions. When defects are present, these interactions are
described by the Hamiltonian:
\be\label{ham-xy}
\frac{\cal{H}}{kT}
=-K\sum_{\langle i j\rangle}{\epsilon_i\epsilon_j\vec{\phi}_i\cdot\vec{\phi}_j},
\ee
where $K=J/kT$ is the coupling (inverse temperature, with $k=J=1$), and
$\epsilon_i=1$ if a spin occupies the site, or $\epsilon_i=0$ otherwise.
Depending on the internal dimension ($m$) of the order parameter, spins are taken
as $\vec{\phi}_i=\pm 1$ for the 3DIS ($m=1$), and
$\vec{\phi}_i=(\cos\theta_i,\sin\theta_i)$ for the 3DXY ($m=2$) model.
For each independent disorder realization, an aerogel is first built by the DLCA
on-lattice algorithm described above \cite{remiju}, taking each site from the complete
final cluster as a defects. Aerogel pores, {\em i.~e.,~}$N=(1-c)L^3$ empty sites left in
the lattice, are then filled with spins, being $c$ is the concentration of defects.
Monte Carlo sweeps (MCS) are then taken as follows, using Hamiltonian (\ref{ham-xy}):
Wolff cluster update algorithm \cite{wolff89} is implemented, taking one sweep as eight
consecutive cluster flips, to reduce autocorrelations caused by critical slowing down.
We quote here the procedure followed in simulations of the 3DXY model \cite{vasquez03}.
For each disorder realization, thermalization is reached after $2\times 10^4$ sweeps,
while $2\times 10^6$ production sweeps are taken at equilibrium for further statistical
analyses. A suitable number of disorder realizations have been taken, depending on
lattice sizes, which run from $L=16$ ($256$) to $L=64$ ($16$). The method is rather
different in the disordered 3DIS\cite{paredes}, in that shorter time series
($\sim 10^3$MCS) are taken in statistics, increasing the number of realizations at each
size ($\sim 2\times 10^3$).
A detailed analysis on self-averaging of thermodynamical quantities, performed for the
3DIS case, shows a lack of self-averaging in the magnetization, suceptibility and
specific heat, and on the other hand, the  energy is weakly self-averaged, with an exponent
$x\approx 2.58$. In that work \cite{paredes}, disorder distributions of thermodynamic
observables for simulations of the 3DIS in aerogels are shown to be more symetric and
narrower than for RDIS. Thus, good rough estimates could be found with not so much
disorder realizations for the aerogel case. For this reason, although such analysis has
not been performed for results presented here for the 3DXY case, an acceptable agreement
with experimental results on the SF transition of $^4$He is obtained \cite{vasquez03}.

\subsection{Observables}
The magnetization squared (order parameter) is calculated as,
\be
M^2=[|\langle\vec{\phi}\rangle|^2]=[|\sum_i{\vec{\phi}_i}/N|^2],
\ee
where brackets represent canonical ensemble averages, while square brackets express
averages over disorder realizations, for fixed $L$ and $c$. The energy per spin is
calculated as the canonical ensemble average:
\be
{\cal{E}}=[\langle{E}\rangle]=[\langle{H}\rangle/N].
\ee

Moments for the energy per spin are estimated at the simulation coupling $K_o$, as
${\cal{E}}^n=[\langle{E^n}\rangle]$, as well as moments $M^n$ for the magnetization are
estimated for the corresponding energy bins. Logarithmic derivatives of moments
$n=1,2,4$ of magnetization are then calculated using the corresponding $E-M^n$
covariance in the canonical ensemble:
\be\label{logeq}
{\cal{D}}_n=\frac{\partial\ln(M^n)}{\partial K}=
-\frac{\langle M^nE\rangle-\langle E\rangle\langle M^n\rangle}{\langle M^n\rangle}.
\ee

The specific heat is calculated through fluctuations of the energy per spin:
\be\label{spheq}
{\cal{C}}=NK^{-2}\left(\langle E^2\rangle-\langle E\rangle^2\right).
\ee

The helicity modulus $\left[\langle\Upsilon_{\hat{\mu}}\rangle\right]$ has been
determined using the Kubo formula \cite{li90,moon95}, averaged over the three
directions $\hat{\mu}=\hat{x}, \hat{y}, \hat{z}$, assuming that the model and disorder
distributions are isotropic. This quantity is proportional to the superfluid density,
as was derived by Fisher and colleagues \cite{fisher73}, so it is directly related to
the order parameter of the SF transition. We use this relation to show that the
confined system (AEXY) acts in $d=3$ external dimension, {\em i.~e.}, aerogel pores are
effectively three-dimensional \cite{vasquez03}.

These quantities have been stored from MC simulations at each realization of disorder
and, after the application of reweighting techniques, extrapolation of thermodynamical
quantities of interest are made at couplings $K\approx K_o$\cite{swendsen88,newman00}.

\subsection{Finite size scaling: thermal exponents}
\begin{wrapfigure}{i}{0.5\textwidth}
\centerline{\includegraphics[width=0.5\textwidth]{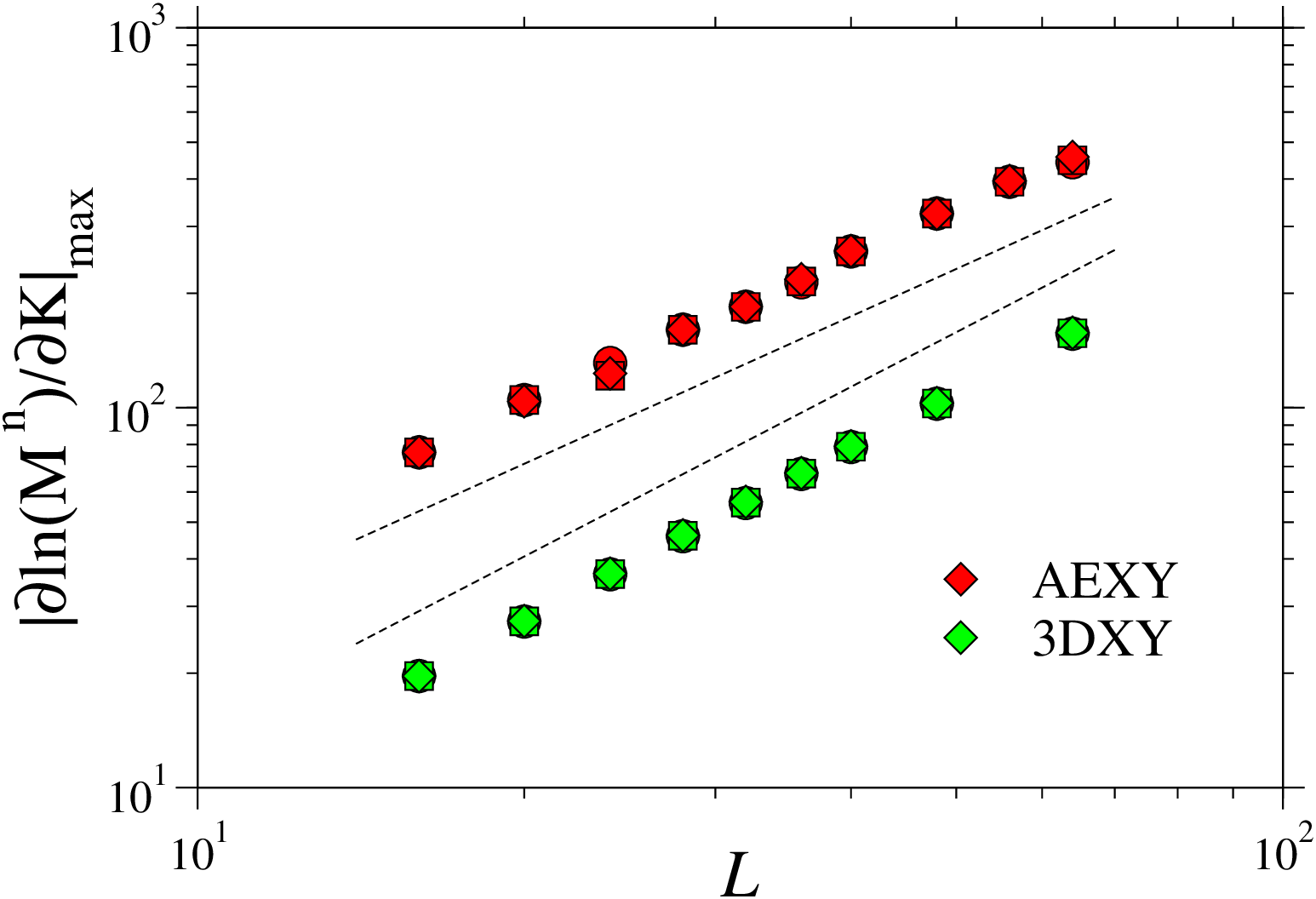}}
\parbox{0.52\textwidth}{
\caption[Correlations in DLCA-GC]
{\label{logplots}{FSS of maxima of logarithmic derivatives ${\cal{D}}_n$ ($n=1,2,4$)
for bulk 3DXY and the AEXY at concentration $c=0.05$ of defects. Corresponding original
plots have been vertically displaced to coincide at $L=16$.
Dashed lines are fits to the FSS expresion (\ref{logfss}) on collected data. }}
}
\end{wrapfigure}
Maxima from reweighted curves for logarithmic derivatives of moments $n=1,2,4$ of the
magnetization (\ref{logeq}), and the specific heat (\ref{spheq}), as functions of the
coupling $K$, are averaged over disorder to estimate thermal critical exponents.

\subsubsection{$O(2)$--vector models: 3DXY and AEXY}\label{xy}
According to the finite-size scaling (FSS) ansatz \cite{fisher72} maxima of logarithmic
derivatives scale with lattice sizes $L$ as:
\be\label{logfss}
\left[{\cal{D}}_n\right]_{max} \sim L^{1/\nu}\,.
\ee
Respectively, data for the 3DXY and the AEXY systems are depicted in Fig.~\ref{logplots},
as function of lattice sizes $L$. Both groups of data have been vertically displaced
(multiplied by a constant) to make corresponding plot coincide at $L=16$, which allows
making a fit on collected data. The FSS power-law fit (\ref{logfss}) for bulk 3DXY, using
lattice sizes $L>24$ gives $1/\nu_{pure}= 1.483(3)$, which agrees well with previously
reported results\cite{hasenbusch99}, and the exponent $\zeta\approx0.67$ reported for bulk
$\lambda$-transition\cite{lipa96}. The same fit to data for the AEXY, at concentration
$c=0.05$, gives $1/\nu_{\mbox{\tiny AEXY}}= 1.29(2)$, clearly different from the bulk
value. Fits on Fig.~\ref{logplots} (dashed lines) have been vertically displaced to
facilitate comparison between the slopes.

The correlation length exponent $\nu_{\mbox{\tiny AEXY}}\approx 0.77$ obtained for this
confined system agrees well with the exponent $\zeta\approx 0.76$ obtained in
experiments\cite{mulders91}. Both results, numerical and experimental, suggest that a LRC
distribution of defects must be affecting the critical behaviour of the 3DXY universality
class, when confined in aerogel-like structures.
After the extended criterion (\ref{whlrc}), this structure must have a decaying exponent
$a < 2/\nu_{pure}\approx 2.98$. Analyses on the DLCA structure resumed above give
$a\approx 2.6$ for gelling clusters of DLCA aerogels at $c=0.05$, so the WH condition is
fulfilled and the distribution GC of disorder could be relevant. Nevertheless, there are
other correlated structures within DLCA aerogels, the backbones, but given that $a> 3$ for
these structures, they do not satisfy the WH condition. Had these structures have a
decaying exponent smaller than $d$, Fig.~\ref{sub-dlca} shows that this exponent is already
greater than the corresponding exponent for the GC.
The analysis resumed by WH\cite{weinrib83} when multiple exponents are present on LRC
strucutures, {\em i.~e.}, when $g(r)=\sum_i{g_ir^{-a_i}}$, it is the term with smallest
exponent which rules the critical behaviour for the impure system.
Indeed, Prudnikov {\em et al.~}\cite{prudnikov00} predict that for a decaying exponent
similar to that reported above for the GC ($a\approx 2.6$), the correlation length
exponent should be $\nu_{\mbox{\tiny LRC}}\approx 0.73$, so both, experimental and
numerical results roughly confirm this prediction. It is worth recalling that the whole
DLCA structure itself is not LRC, but it is a mixture of SRC structures with a clearly LRC
structure, the GC. As follows from WH results, SRC subsets are {\em irrelevant}. Thus, evident
changes observed in critical exponents for this univerality class suggest that
only the LRC subset rules the critical behaviour of both systems, SF $^4$He and 3DXY
interacting spins.
\begin{wrapfigure}{i}{0.5\textwidth}
\vspace{0.2cm}
\centerline{\includegraphics[width=0.5\textwidth]{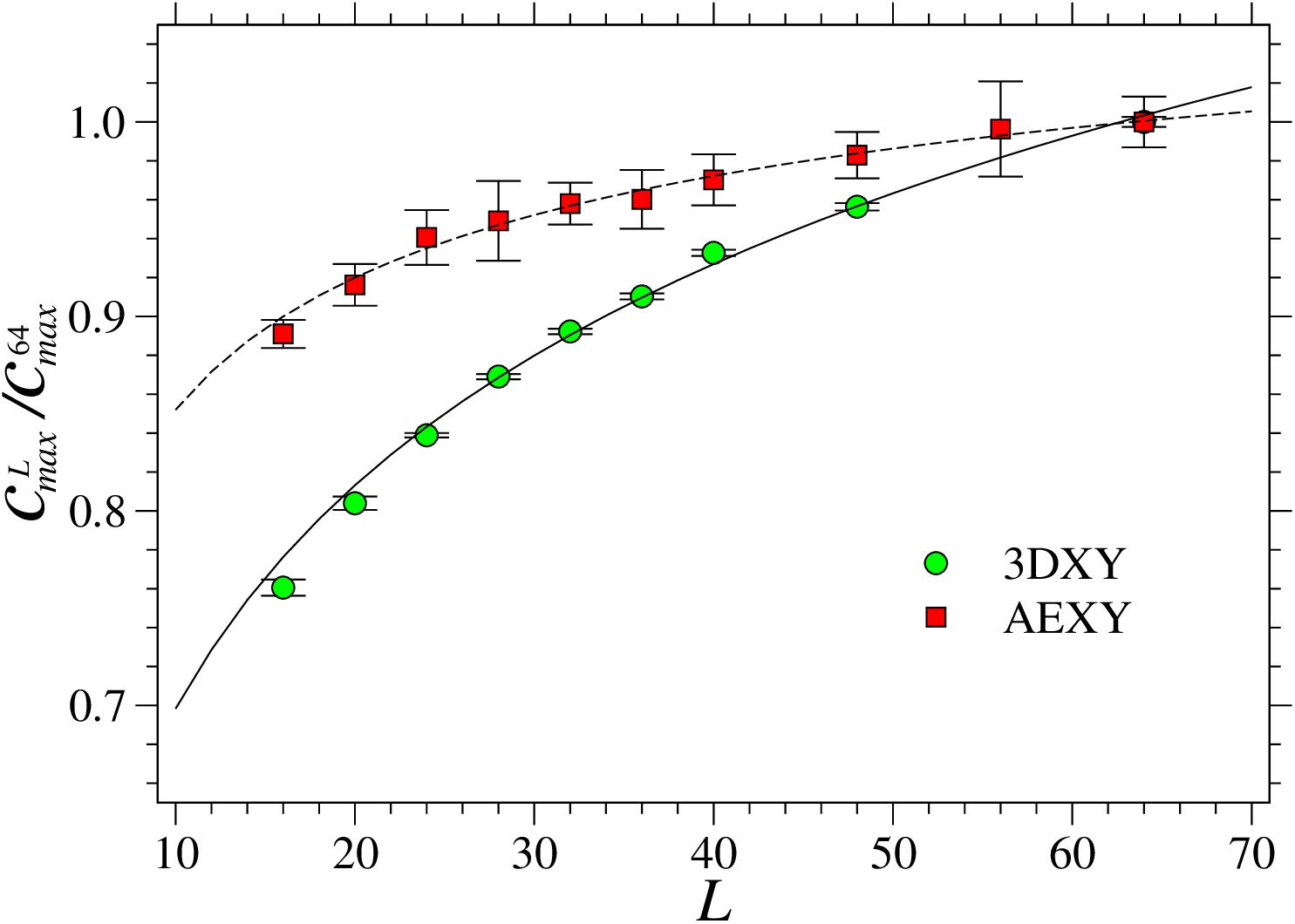}}
\centerline{\parbox{0.55\textwidth}{
\caption[Correlations in DLCA-GC]
{\label{sphplots}{FSS of maxima of the specific heat for bulk 3DXY (circles)  and
confined AEXY at concentration $c=0.05$ of defects (squares). Lines are fits to the
FSS expresion (\ref{sphfss}). }}
}}
\end{wrapfigure}

In addition, we have further determined critical exponents for the specific heat and the
helicity modulus as well. Maxima for the specific heat, assuming that $\alpha<0$, scale as:
\be\label{sphfss}
\left[{\cal{C}}\right]_{max}\sim {\cal{C}}_{\infty}+AL^{\alpha/\nu}\,.
\ee
These maxima have been depicted in Fig.~\ref{sphplots} as function of $L$, normalized
to $c(L=64)\equiv 1$, for bulk 3DXY (circles) and the confined AEXY system at $c=0.05$
(squares). Non-linear fits of corresponding points to the FSS expresion (\ref{sphfss}),
using data for $L\geq 28$ to avoid corrections to scaling, are shown in
Fig.~\ref{sphplots} by corresponding lines. The method was developed by Schultka and
Manousakis\cite{manousakis95}, to determine the specific heat exponent for the pure
3DXY system. The corresponding best fits give $\alpha/\nu\approx -0.015$ for the pure
system, and $\alpha/\nu\approx -0.38$ for the confined system. These results agree well
with our results on correlation length exponents, shown above, after the Josephson
hyperscaling relation \cite{joseph66} in the form $2/\nu-\alpha/\nu=d^*$, which gives
$d^*\approx 3.0$ for both systems. Using corresponding results on the correlation length
exponents, we obtain $\alpha\approx -0.010$ for bulk 3DXY, which agrees well with
calorimetric studies on SF $^4$He\cite{lipa96}. The result $\alpha=-0.29$ for the AEXY is
lower than the reported experimental value $\alpha-0.57$ for $^4$He confined in aerogel
pores at the same volume fraction $c=0.05$ \cite{yoon98}. However, in numerical simulations
reported here, the question posed by authors about the apparent violation of hyperscaling
seems not to emerge from our results on exponents $\nu$ and $\alpha$.
\begin{wrapfigure}{i}{0.5\textwidth}
\vspace{0.2cm}
\centerline{\includegraphics[width=0.5\textwidth]{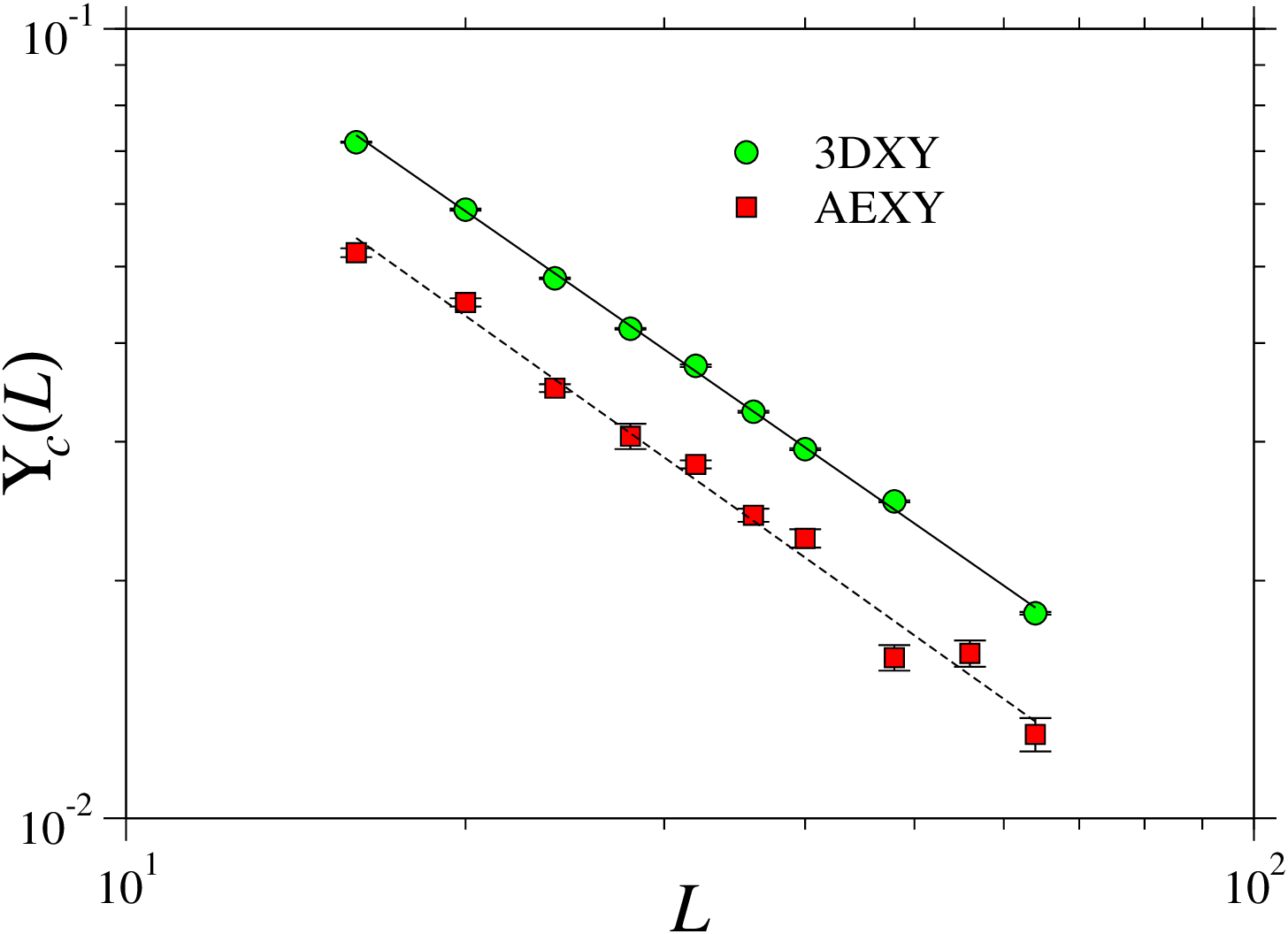}}
\parbox{0.52\textwidth}{
\caption[Helicity modulus]
{\label{upsilon}
FSS plots of the helicity modulus at $K_c$. Bulk 3DXY (circles) and confined AEXY
at concentration $c=0.05$ of defects (squares). }
}
\end{wrapfigure}
The critical coupling $K_c$ has been determined using the crossing of reweighted plots
of the Binder fourth cumulant for the magnetization
$U_4=1-{\langle M^4\rangle/3\langle M^2\rangle^2}$ at different lattice sizes $L$,
which is universal at $K_c$ (independent on $L$)\cite{binder81}. This analysis yields
$K_c\approx 0.45416$ for the pure system, while the crossing of average reweighted
curves in the AEXY case gives $K_c\approx 0.46495$.

Using these values, we estimate the helicity modulus at $K_c$, averaged over disorder
realizations:
\be\label{upsfss}
\left[\langle\Upsilon_{\hat{\mu}}\rangle\right]\sim L^{-\upsilon/\nu}\,.
\ee
Fig.~\ref{upsilon} shows corresponding data for the helicity modulus in the pure 3DXY
case (circles), and the AEXY case (squares). Lines represent fits to the FSS expression
\ref{upsfss}. The exponent $\upsilon/\nu=1$ confirms that the effective external
dimension for the system is $d=3$, through the hyperuniversality relation
$\upsilon = (d-2)\nu$ \cite{fisher73}. This result also confirms that the correlation
length exponent we have estimated here corresponds to the exponent $\zeta$ for the SF
transition of $^4$He in aerogels.

Results reported here seem to answer the question quoted above from
Refs.~\cite{chan88,chan96}: numerical evidence has been presented here (Section \ref{dlca})
that a physically well defined {\em hidden} LRC distribution of defects exists in aerogels,
which explain changes on the critical behaviour of SF $^4$He when confined in those
structures.

\subsubsection{$O(1)$--vector models: 3DIS and AEIS}
\begin{wrapfigure}{i}{0.5\textwidth}
\vspace{0.2cm}
\centerline{\includegraphics[width=0.5\textwidth]{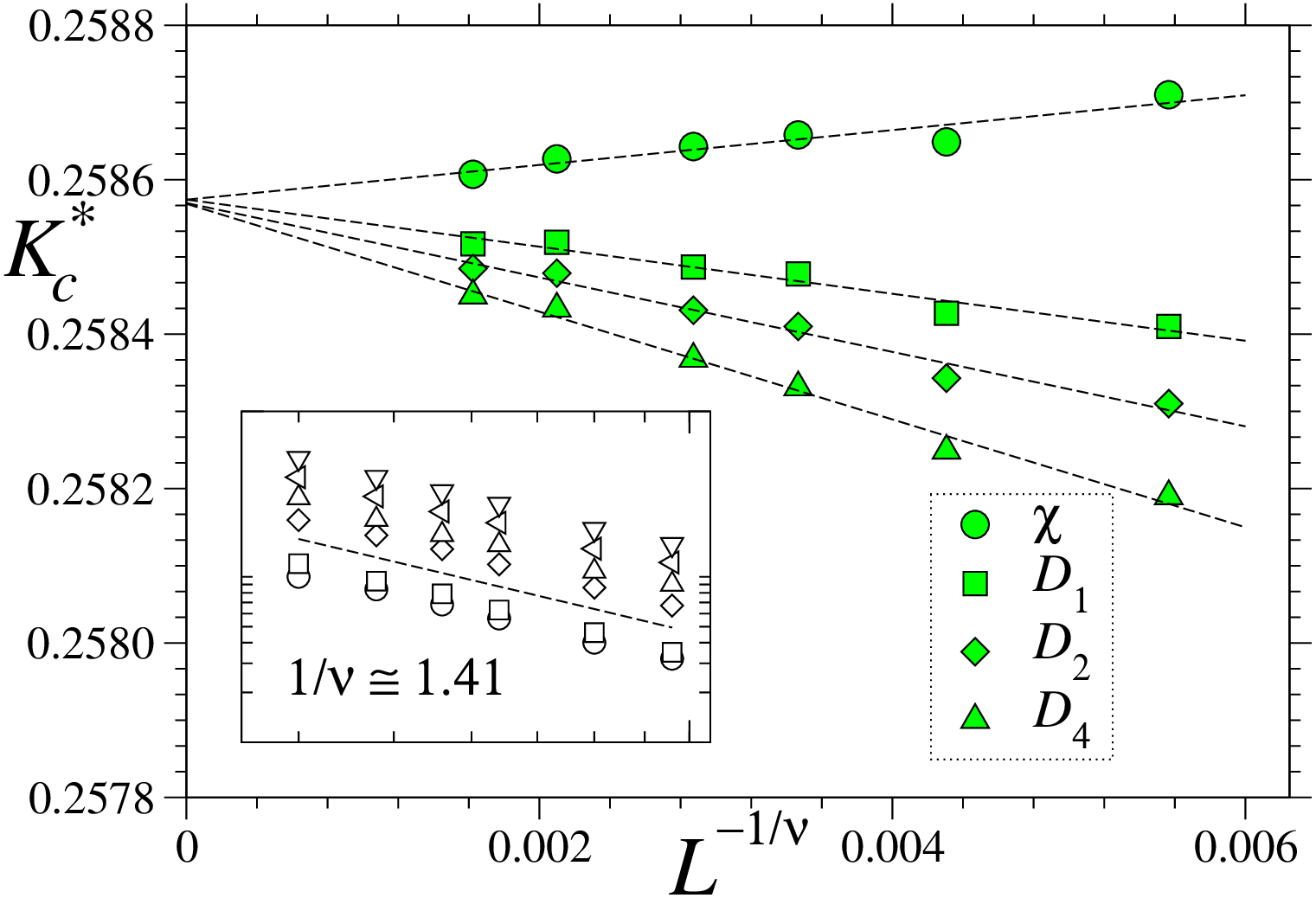}}
\parbox{0.52\textwidth}{
\caption[AEIS critical coupling]
{\label{aeisKc}
FSS plots of pseudocritical couplings $K^*_c$ for the AEIS at $c=0.2$, taken from
positions of maxima for the susceptibility ($\chi$), and logarithmic derivatives
(${\cal{D}}_n$). The horizontal axis have been rescaled to $L^{-1/\nu}$, taking
$1/\nu=1.41$ (estimated in the {\em insert}). }
}
\end{wrapfigure}

We turn now to a brief discussion on changes of the correlation length exponent for the
3DIS, when confined in DLCA aerogels at concentration $c=0.2$ of defects. A detailed
analysis is reported elsewhere\cite{paredes}. The method, as stated above, is quite
different from that used in the AEXY case in that shorter MC time-series are used for
extrapolation, but total numbers of realizations have been incremented in two orders of
magnitude. Shorter CPU times, in this case, allows also making simulations at greater
lattice sizes, being $L=8-96$ in this case. After extrapolation by reweighting, curves
have been averaged over disorder for each $K$, so a unique curve is obtained for each
set of data. Position of maxima ({\em pseudocritical couplings} $K^*_c(L)$) for the magnetic
susceptibility ($\chi$), and logarithmic derivatives ${\cal{D}}_n$ of moments of the
magnetization ($n=1,2,4$) have been estimated through the averaged curves. According to
the finite size scaling theory \cite{binder81}, maxima for a given observable
${\cal{O}}$ scale as
\be\label{aeisfss}
K^*_c({\cal{O}},L)=K_c+a({\cal{O}})L^{-1/\nu} + \mbox{c.~s.}
\ee
being $K_c$ the critical coupling at the thermodynamical limit, $L\to\!\infty$.
Coefficients $a({\cal{O}})$ are non-universal, {\em i.~e.~}, dependent on the observable
${\cal{O}}$ and details of the system. Corrections to scaling (c.~s.~) are avoided
in the analysis reported below, taking data for large enough lattice sizes ($L>40$).
Points depicted on the inset of Fig.~\ref{aeisKc}, have been calculated through local
differences $K^*_c({\cal{O}}_i,L)-K^*_c({\cal{O}}_j,L)$ between data from two different
observables ${\cal{O}}_i$ and ${\cal{O}}_j$. This transformation of data allows
eliminating the non-singular term ($K_c$) from Eq.~(\ref{aeisfss}), common by hypothesis
to all observables, allowing an easier estimation of the exponent on the leading term
$aL^{-1/\nu}$.

Power-law fits have been made to these transformed data, giving a rough
estimate of the (common) FSS exponent $1/\nu\approx 1.41$. This approximate has been
used to rescale the horizontal axis to $L^{-1/\nu}$, as indicated in Fig.~\ref{aeisKc}.
As expected, a linear behaviour is observed for all data, and linear fits to the
expression (\ref{aeisfss}) gives just one value $K_c = 0.25857(1)$ for all plots.
A more detailed analysis on the scaling of logarithmic derivatives at the critical
point reveals, however, that the correlation length exponent  tends to a value  closer to
$3/2$\cite{paredes}.
The correlation length exponent $\nu_{AEIS}\gtrsim 0.67$ obtained is clearly greater than
that for the pure 3DIS and the LV critical point ($\nu\approx 0.63$). These results point
to a negative or zero value for the specific heat exponent $\alpha$. Experiments on the
LV critical point of $^4$He in aerogels\cite{awong90} report specific heat curves that seem
to have a finite peak at $T_c$, but authors do not estimate the critical exponent
$\alpha$ which may either be negative or zero as well. After Josephson hyperscaling,
this is consistent with our rough result $1/\nu\lesssim 1.5$, tending to values
closer to that predicted for the SRC Ising universality class than that predicted for the
LRC fixed point\cite{prudnikov00}.

We must then conclude, in this case, that the FSS method applied here may not be precise
enough to determine really asymptotic values for critical exponents, but a remarkable
difference exist between results approached here and exponents for the pure system, as
expected after Harris criterion.

\section{Concluding remarks}
We have reported here on the effects of a specific type of disordered distributions of
defects on the critical behaviour of $O(m)$--vector models, using the $O(1)$ and the
$O(2)$ models, represented by the 3DIS and the 3DXY models, respectively. A detailed
scaling analysis has been made to structures created through the DLCA on-lattice
algorithm, which reproduces well the geometrical tasks of real silica aerogels. This
analysis shows that the whole structure is not, strictly speaking, a fractal:
the power-law decaying regime is preserved only through a few lattice constants, and
even more, it does not scale with lattice sizes. A more detailed dissection, shows the
existence of a physically well defined subsets within the whole strucuture, gelling
clusters, that keep the algebraic decaying regime through lengths comparable to the
lattice size, with a cutoff that scales with the system size. Under the light of WH
theory, the existence of this fractal (LRC) subset already explains changes on critical
exponents of the 3DXY model, when confined in these aerogel-like simulated structures,
respect to those of the pure system. Results for the confined (AEXY) system roughly
confirm those observed for the SF transition of $^4$He inside aerogels, and explain
why a crossover to bulk values is not observed in critical exponents reported from
experiments, as $T\to T_c$, when the correlation length diverges\cite{yoon98}.
Even more intriguing are results from the 3DIS confined in the same type of disorder
distribution, as exponents seem to be affected by the LRC fixed point to finally tend
values far from those predicted by theory\cite{prudnikov00}, and closer to those of
the SRC fixed point. The question about the critical behaviour of the AEIS rests open
as conclusive results on asymptotic values for critical exponents have not yet been
obtained. Previous renormalization group flows to fixed points, sketched by Weinrib
and Halperin\cite{weinrib83}, and analyses made by Prudnikov {\em et al.~}
\cite{prudnikov00}, point out that the approach to the LRC fixed point may be
oscillating, which can make very dificult to enter a really asymptotic regime applying
conventional finite size scaling.

\section*{Acknowledgements}
Authors thank CNRS and FONACIT (PI2004000007) for their support. Invaluable discussions
with A.~Hasmy and R.~Jullien have improved our understanding on aerogel structure.
C.~V\'asquez kindly acknowledges the collaboration and technical support received by the
personnel of the LCVN at Montpellier, France.

\newpage
\section*{Questions and answers}

$\boldsymbol{\cal Q}$ ({\em Rafael Rangel}): {\bf What are the theoretical predictions for the (possible)
        change of exponents by disorder?}

$\boldsymbol{\cal A:}$ Harris criterion predicts that a weak uncorrelated disorder is not
relevant if the correlation length exponent $\nu$ is larger  than $2/d$ for the pure
system, or after Josephson hyperscaling, $\alpha=2-d\nu<0$. For 3DIS case, $\alpha > 0$,
and then  short-range-correlated disorder could be relevant. Weinrib and Halperin
extend this criterion to cases where there exist long-range correlations (LRC) in the
distribution of defects, as expressed by Eqs.~(\ref{whsrc}) and (\ref{whlrc}). It is
this extension that allows explaining changes on the critical behaviour of the 3DXY
model when confined in aerogel-like distributions of defects, and changes on critical
exponents of the SF transition of $^4$He in aerogels as well.

$\boldsymbol{\cal Q}$ ({\em Dragi Karevski}): {\bf How does your FSS technique work if there are different
correlations?}

$\boldsymbol{\cal A:}$ Weinrib and Halperin \cite{weinrib83} already considered the case
where there exist terms with different decaying exponents in the correlation function
$g(r)$, as was quoted in Subsection \ref{xy}: the critical behaviour couples to the term
with the lowest exponent. Indeed, in the model disorder we are considering in simulations
presented here, DLCA, there exist several subsets of the whole cluster which present
different decaying regimes. Results point to state that, in the 3DXY confined case
(AEXY), it is the gelling cluster (GC) that rules the critical behaviour, as it is the
LRC subset with the lowest exponent (see Fig.~\ref{sub-dlca}). The AEIS case is by far
more interesting in that, competing effects are present, as SRC disorder is also
relevant. Being the LRC fixed point marginal, it seems that it is the SRC fixed point that
finally governs the critical behaviour, {\em i.~e.}, the effective exponent $1/\nu$ turns
from being attracted to $1/\nu\approx 1.4$ (predicted by Prudnikov {\em et al.~}
\cite{prudnikov00} for the LRC fixed point) to $1/\nu\lesssim 3/2$, closer to the SRC
fixed point value. Nevertheless, even if only one model LRC distribution of defects
were present, the application of FSS techniques must consider that the convergence of
the correlation function to the final decaying regime $L^{-(d-d_f)}$ will depend on the
lattice size $L$, so the convergence of {\em effective exponents} to critical values will be
ruled by an exponent that asymptotically tends to a stable value. This task of the
correlation function may cause succesive crossover regimes which appear in the convergence
of effective exponents to asymptotic critical exponents.

$\boldsymbol{\cal Q}$ ({\em Dragi Karevski}): {\bf How does the FSS with $V=L^d$ (and not with $L$) hold?}

$\boldsymbol{\cal A:}$ The FSS theory is based on the homogeneity of the free energy
density, which allows writing it in terms of the {\em scaling variable} $L/\xi=Lt^{\nu}$,
which at its time may be redimensioned to $tL^{-1/\nu}$. The external dimension $d$ of the
system appears only in corresponding hyperscaling relations, such as those of Josephson
$2/\nu-\alpha/\nu=d$ and Rushbrooke $\gamma/\nu+2\beta/\nu=d$.

$\boldsymbol{\cal Q}$ ({\em Wolfhard Janke}): {\bf What is the dimension $d_f$ of your fractal?}

$\boldsymbol{\cal A:}$ The whole DLCA clusters in our model disorder are not fractal,
as they enter homogeneous regime at very few lattice constants, but the well defined GC
have correlations that seem to span all scales at the thermodynamical limit. The best
estimate we have of the fractal dimension for these objects is $d_f\simeq 1.7$.

$\boldsymbol{\cal Q}$ ({\em Yurii Holovatch}): {\bf How is $c$ related to $a$?}

$\boldsymbol{\cal A:}$ As the concentration of defects increases, the fractal dimension of
the GC increases as well, as stated in Ref.~\cite{hasmy94}, so the exponent $a$ decreases.

%
%
  \label{last@page}

\begin{thebibliography}{7}
\bibitem{comment}
This is not, of course, a rigurous mathematical prove. Authors in
Refs.~\cite{weinrib83,prudnikov00} define $g(r)$ as the correlation function for the local
critical temperature $g(|\vec{\boldsymbol{r}}_1-\vec{\boldsymbol{r}_2}|)=\langle T_c(\vec{\boldsymbol{r}}_1)T_c(\vec{\boldsymbol{r}}_2)\rangle\sim |\vec{\boldsymbol{r}}_1-\vec{\boldsymbol{r}_2}|^{-a}$. We relate fluctuations in this local $T_c(\vec{\boldsymbol{r}})$ with the
local density of surrounding defects $n(\vec{\boldsymbol{r}})$, and it is in terms of this
density function that we made approximate calculations about $g(r)$, which allow relating
the exponent $a$ to the fractal dimension $d_f$ of the distribution.
\end{thebibliography}
  \end{document}